\documentclass[a4paper,american,aps,prl,twocolumn,utf8,floatfix]{revtex4}
\usepackage[T1]{fontenc}
\usepackage[utf8]{inputenc}
\usepackage{prettyref}
\usepackage{amsmath}
\usepackage{amssymb}
\usepackage{graphicx}

\makeatletter

\pdfpageheight\paperheight
\pdfpagewidth\paperwidth

\newcommand{\lyxdot}{.}

\@ifundefined{textcolor}{}
{%
 \definecolor{BLACK}{gray}{0}
 \definecolor{WHITE}{gray}{1}
 \definecolor{RED}{rgb}{1,0,0}
 \definecolor{GREEN}{rgb}{0,1,0}
 \definecolor{BLUE}{rgb}{0,0,1}
 \definecolor{CYAN}{cmyk}{1,0,0,0}
 \definecolor{MAGENTA}{cmyk}{0,1,0,0}
 \definecolor{YELLOW}{cmyk}{0,0,1,0}
}

\usepackage{hyperref}
\usepackage{natbib}
\makeatother

\usepackage{babel}
\begin{document}

\preprint{This line only printed with preprint option}

\title{A thermodynamically self-consistent non-stochastic micromagnetic
model for the ferromagnetic state}

\author{Mykola Dvornik}

\email{Mykola.Dvornik@ugent.be}

\homepage{http://dynamat.ugent.be}

\selectlanguage{american}%

\affiliation{DyNaMat Lab, Ghent University, Krijgslaan 281/S1, 9000 Ghent, Belgium}

\author{Arne Vansteenkiste}

\affiliation{DyNaMat Lab, Ghent University, Krijgslaan 281/S1, 9000 Ghent, Belgium}

\author{Bartel Van Waeyenberge}

\affiliation{DyNaMat Lab, Ghent University, Krijgslaan 281/S1, 9000 Ghent, Belgium}
\begin{abstract}
In this work, a self-consistent thermodynamic approach to micromagnetism
is presented. The magnetic degrees of freedom are modeled using the
Landau-Lifshitz-Baryakhtar theory, that separates the different contributions
to the magnetic damping, and thereby allows them to be coupled to
the electron and phonon systems in a self-consistent way. We show
that this model can quantitatively reproduce ultrafast magnetization
dynamics in Nickel. 
\end{abstract}
\maketitle
Micromagnetism typically relies on the phenomenological theory developed
by Landau and Lifshitz\citep{Landau1935} that is used to fit experimental
data on a wide range of magnetization dynamics problems. However,
this theory cannot be applied to ultrafast magnetization phenomena
and spintronics, since it cannot account for changes of the magnetization
vector length (so-called longitudinal magnetization dynamics)\citep{Beaurepaire1996,Koopmans2005,Radu2009,Koopmans2010}
and non-local exchange damping\citep{Zhang2009,Nembach2011,Faehnle2013,Turgut2013}.
The Landau-Lifshitz approach unifies all possible relaxation mechanisms
into one isotropic relaxation term of relativistic nature, making
different contributions to the intrinsic damping indistinguishable.
However, as we approach the timescale of the exchange interaction,
the difference in nature, symmetry properties and strength of the
spin-electron and spin-phonon relaxation channels become more prominent.
So the Landau-Lifshitz approach is only valid for qubic lattices and
in case of relatively slow transverse magnetization dynamics in a
temperature range far away from the Curie point. As we already mentioned,
many emerging topics of the nanomagnetism lie beyond this regime.

An advance has been made by \citet{Atxitia2010} and \citet{Atxitia2011}
by using a combination of the two-temperature (2T) model \citep{Kaganov1957}
and the Landau-Lifshitz-Bloch (LLB) equation \citep{Garanin1997,Mayergoyz2012}
to describe magnetic response at elevated temperatures. This approach
naturally accounts for the longitudinal magnetization dynamics, but
still relies on a single relativistic scalar coupling-to-the-bath
parameter and, thereby, does not separately describe the relaxations
of different origin, nor it includes the non-local damping. Secondly,
the model neglects spin-electron and spin-phonon energy flows, thereby
violating the energy conservation law (i.e. the model uses the heat
bath approximation for both electron and phonon systems with respect
to the spins). The same approach is employed by \citet{Koopmans2010}
in the so-called M3T model. This model is specifically developed to
explain ultrafast thermal demagnetization, but neglects precessional
dynamics as second-order effect (on a much larger timescale). It has
been demonstrated, that both models are equivalent for the description
of the longitudinal dynamics\citep{Atxitia2011}.

At the same time self-consistent atomistic models have been reported
by \citet{Ma2012} and \citet{Chimata2012}. Although these theories
comply with the energy conservation law, they (a) oversimplify the
temperature dependence of the heat capacity of spins and the spin-electron
coupling constant and (b) cannot be easily mapped onto the experimentally
accessible micromagnetic scale. So there is no universal approach
to finite temperature micromagnetism. 

In this letter we report on the development of a micromagnetic model
that unifies and extends the previously developed approaches within
a single framework. Although the model is phenomenological, it allows
for a consistent physical interpretation of the spin relaxation terms
and their coupling to electron and phonon systems.

We rely on Baryakhtar's theory \citep{Baryakhtar1986,Baryakhtar1998,Baryakhtar2010}
that derives the following equation for the magnetization dynamics
and relaxation:

\begin{equation}
\frac{\mathrm{\partial\mathbf{M}}}{\mathbf{\mathrm{\partial}\mathrm{t}}}=-\gamma_{LL}\mathbf{M}\times\mathbf{H}+\mathbf{\hat{\lambda}}(\mathbf{M})\mathbf{H}-\mathbf{\hat{\lambda}_{\mathbf{\mathit{pq}}}^{\mathrm{(e)}}}\frac{\partial^{2}\mathbf{H}}{\partial x_{p}\partial x_{q}}\label{eq:LLBr-general}
\end{equation}
where $\mathbf{M}$, $\mathbf{\gamma}$, $\mathbf{H}$ are the magnetization
vector, the positively defined gyromagnetic ratio and the internal
effective field, respectively. The terms in the right-hand side of
the equation describe the magnetic torque, the local relativistic
and the non-local exchange relaxations, respectively. The LLB equation
is the special case of eq. \eqref{eq:LLBr-general}.

The internal field is given by $\mathbf{H=\mathrm{-}\frac{\delta\mathit{w}}{\delta M}}=\mathbf{H}^{MM}+\mathbf{H}^{\parallel}$,
where $\mathit{w}$ is the Gibbs free energy density of the magnetic
medium, $\mathbf{H}^{MM}$ and $\mathbf{H}^{\parallel}$ denote micromagnetic
and longitudinal effective fields. The longitudinal field (historically
referred to as ``molecular'' field) is the micromagnetic representation
of the microscopic exchange field that arises from interaction of
the spins with other quasiparticles, namely electrons and phonons.
It drives the spin system to thermodynamic equilibrium via angular
momentum transfer. This field consists of two parts $\mathbf{H}^{\parallel}=\mathbf{H}_{s-e}^{\parallel}+\mathbf{H}_{s-p}^{\parallel}$,
i.e. a spin-electron ($s-e$) and a spin-phonon ($s-p$) contribution\citep{Sultan2012}.
The field implicitly accounts for the spin fluctuations and optical
magnons that cannot be resolved on the micromagnetic scale. It has
the following phenomenological form\citep{Garanin1997}
\begin{gather}
\mathbf{H}_{i}^{\parallel}=\frac{nk_{B}T_{i}}{\mu_{0}M_{s}}\frac{B_{J}(m\frac{\theta}{T_{i}})-m}{B_{J}^{'}(m\frac{\theta}{T_{i}})}\mathbf{\frac{m}{\mathbf{\mathit{m}}}}\label{eq:longfield}
\end{gather}
where $T_{i}$ is the temperature of the corresponding quasi-particles
($i=s-e,s-p$). $n$, $M_{s}$ and $J$ are the number density of
atoms, zero-temperature saturation magnetization and atomic moment,
respectively and $\mathbf{m=M}/M_{s}$. $\theta=\frac{3J^{2}}{J(J+1)}T_{c}$,
where $T_{c}$ is the Curie temperature of the material. $B_{J}$
denotes the Brillouin function for the given atomic moment. Eq. \eqref{eq:longfield}
accounts only for competition between the microscopic exchange and
thermal fluctuations energies. The contribution of the Zeeman interaction
is explicitly included into the model. Although eq. \eqref{eq:LLBr-general}
and eq. \eqref{eq:longfield} successfully describe the ferromagnetic-to-paramagnetic
phase transition, the opposite transition cannot be modeled without
explicitly considering short-order spin fluctuations\citep{Oguchi1955},
e.g. by introducing a stochastic form of the model. This regime lies
beyond the scope of the present study.

The rank-2 tensors $\mathbf{\hat{\lambda}}$ and $\mathbf{\hat{\lambda}_{\mathbf{\mathit{pq}}}^{\mathrm{(e)}}}$
describe the strength of the relativistic and exchange dissipations,
respectively. In contrast to the Landau-Lifshitz and Landau-Lifshitz-Bloch
models, the relativistic relaxation tensor obeys the crystallographic
and magnetic symmetries of the system. So let us expand it into powers
of $\mathbf{M}$ around the highest symmetry magnetic state $M=0$\citep{Baryakhtar2010},
i. e.

\begin{gather}
\mathbf{\hat{\lambda}}(\mathbf{M})=\mathbf{\hat{\lambda}_{\mathit{\mathrm{\parallel}}}}+\hat{\mu}{}_{pq}M_{p}M_{q}+O(M^{4})\label{eq:tensor-expansion}
\end{gather}
The relativistic relaxation tensors $\hat{\lambda}_{\parallel}$ and
$\hat{\mu}$ mimic the crystallographic symmetry of the system over
the corresponding spatial indices. Substituting of eq. \prettyref{eq:tensor-expansion}
into eq. \prettyref{eq:LLBr-general} and assuming at least a uniaxial
symmetry of the media leads to

\begin{multline}
\frac{\mathrm{\partial\mathbf{M}}}{\mathbf{\mathrm{\partial}\mathrm{t}}}=-\gamma_{LL}\mathbf{M}\times\mathbf{H}+\\
+\hat{\lambda}_{\mathrm{\parallel}}\mathbf{H}-\mathbf{\hat{\lambda}}^{(e)}\Delta\mathbf{H}+\\
+\hat{\mu}_{\parallel}(\mathbf{M}\cdot\mathbf{H})\cdot\mathbf{M}-\mathbf{M}\times\hat{\mu}_{\perp}(\mathbf{M}\times\mathbf{H})\label{eq:LLBr-expanded}
\end{multline}
where $\hat{\mu}_{\perp}$ and $\hat{\mu}_{\parallel}$ are decomposed
from $\hat{\mu}{}_{pq}$ to separate conservative and non-conservative
second-order relaxations, respectively. Let us rewrite all the relaxation
tensors via dimensionless coupling constants

\begin{gather}
\mathbf{\hat{\lambda}}_{\parallel}=-\gamma_{LL}M_{s}\alpha_{\parallel}\hat{\nu}\nonumber \\
\hat{\lambda}^{(e)}=-\gamma_{LL}M_{s}a^{2}\alpha_{\parallel}^{(e)}\hat{\nu}\nonumber \\
\hat{\mu}_{\perp}=-\frac{\gamma_{LL}\mu_{\perp}}{M_{s}}\hat{\nu}\label{eq:dimensionless}\\
\hat{\mu}_{\parallel}=-\frac{\gamma_{LL}\mu_{\parallel}}{M_{s}}\hat{\nu}\nonumber 
\end{gather}
where $a$ is the lattice constant, $\hat{\nu}$ describes crystallographic
symmetry of the media, while $\alpha_{\parallel}$, $\alpha_{\parallel}^{(e)}$,
$\mu_{\perp}$and $\mu_{\parallel}$ define strength of the corresponding
relaxations. Substitution of \eqref{eq:dimensionless} into \eqref{eq:LLBr-expanded}
we finally arrive at 

\begin{align}
\frac{1}{\gamma_{LL}}\frac{\mathrm{\partial\mathbf{m}}}{\mathbf{\mathrm{\partial}\mathrm{t}}} & =-\mathbf{m}\times\mathbf{H}+\mathbf{R_{\mathit{s-e}}^{\mathit{}}}+\mathbf{R_{\mathit{s-p}}}\nonumber \\
\mathbf{R}_{\mathit{s-e}}^{\mathit{}} & =\alpha_{\parallel}\hat{\nu}\mathbf{H}-\alpha_{\parallel}^{(e)}a^{2}\hat{\nu}\Delta\mathbf{H}\nonumber \\
\mathbf{R}_{\mathit{s-p}}^{\mathit{}} & =\mu_{\parallel}\hat{\nu}(\mathbf{m}\mathbf{H})\mathbf{m}-\mathbf{\mu_{\perp}}\mbox{\ensuremath{\mathbf{m}}}\times\hat{\nu}(\mathbf{m}\times\mathbf{H})\label{eq:LLBar}
\end{align}
The important property of the expansion \eqref{eq:tensor-expansion}
is that it separates the zeroth-order paramagnetic relaxation $\mathbf{R}_{s-e}$
(independent of $\mathbf{m}$) from the higher-order magnonic relaxation
$\mathbf{R}_{s-p}$ (that depends on $\mathbf{m}$). If we neglect
the non-local relaxation and assume that the length of the magnetization
vector is conserved, then eq. \eqref{eq:LLBar} reduces to the Landau-Lifshitz
equation with a damping constant that is the sum of Baryakhtar's local
relaxation constants $\alpha_{LL}=\alpha_{\parallel}+\mu_{\perp}$.

Baryakhtar has pointed out that the expansion of the relaxation tensors
\eqref{eq:tensor-expansion} is analogous to the expansion of the
Gibbs free energy of a magnetic medium. So the quadratic relaxation
terms in eq. \eqref{eq:LLBar} should describe relaxation due to the
spin-orbit and, thereby, spin-phonon coupling\citep{Sparks1961,Pincus1961,A.G.Gurevich1996}.
This conclusion is also consistent with the spin-electron-lattice
model developed by \citet{Ma2012} who showed the spin-phonon coupling
is at least a four-spin correlation function. At the same time the
zeroth-order relativistic paramagnetic relaxation is independent of
the magnetic configuration, and so cannot be attributed to spin-orbit
coupling. According to \citet{Overhauser1953}, the main relativistic
contribution to the relaxation in paramagnets is the spin-electron
spin-flip scattering (due to the interactions with the electron spin
and current) . Finally, according to the review of \citet{Faehnle2013},
the zero-oder non-local damping in metals of a form similar to eq.
\eqref{eq:tensor-expansion} is due to spin-electron \textit{s-d}
exchange interaction \citep{Heine1967}. So to summarize, the zeroth-order
and higher-order relaxation terms in eq. \eqref{eq:LLBar} must describe
spin-electron and spin-phonon couplings, respectively. This classification
is the first important feature of the proposed micromagnetic model.

The second problem addressed by this model is the compliance with
the energy conservation law. To couple the LLBar model to the corresponding
heat transfer equations, we calculate the rate of the energy density
change due to the different dissipations. We emphasize that the Gibbs
free energy depends on the entropy of the system, i.e. $w=w(s),$
where $s$ is the entropy density. So the rate of the energy density
change at the constant entropy density is given by the following expression

\begin{equation}
(\frac{dw_{i}}{dt})_{s}=\frac{\partial w_{i}}{\partial t}-\frac{\partial w_{i}}{\partial s}\frac{\partial s}{\partial t}=\frac{\partial w_{i}}{\partial t}-2\frac{\partial q_{i}}{\partial t}\label{eq:energy-dissipation-rate}
\end{equation}
where $\frac{\partial w_{i}}{\partial t}=-\gamma_{LL}nk_{B}\theta\mathbf{mR}_{i}$
describes change of the internal energy density due to the the molecular
field \citep{Majlis2007} and $\frac{\partial q_{i}}{\partial t}=\frac{1}{2}\frac{\partial w_{i}}{\partial s}\frac{\partial s}{\partial T}=\frac{1}{2}\gamma\mu_{0}M_{s}\mathbf{HR}_{i}$
is the Baryakhtar's dissipative function. The indices denote the type
of the relaxation channel. Finally, we assume that the thermodynamics
of the electrons and phonons is governed by a 2T model

\begin{gather}
C_{e}(T_{e})\frac{\partial T_{e}}{\partial t}=k_{e}\Delta T_{e}+G_{e-p}(T_{p}-T_{e})-\frac{dw_{s-e}}{dt}\nonumber \\
C_{p}(T_{p})\frac{\partial T_{p}}{\partial t}=k_{p}\Delta T_{p}+G_{e-p}(T_{e}-T_{p})-\frac{dw_{s-p}}{dt}\label{eq:2TM}
\end{gather}
where $T_{i}$, $C_{i}$, $k_{i}$ and $G_{e-p}$ are the temperature,
volume-specific heat capacity, heat conductance and macroscopic electron-phonon
coupling constant, respectively. The volume-specific heat capacity
of the phonons is estimated from the Debye model, while for the electrons
a linear temperature law is assumed $C_{e}=\gamma T_{e}$. In general,
the electron-phonon coupling $G_{e-p}$ and $\gamma$ both depend
on the temperature of the electrons. For the sake of simplicity we
assume that these quantities are constant. We should be able to refine
our model by calculating them from the Fermi distribution and DOS
of electrons, but this approach lies beyond the scope of the present
study. Finally, the system of coupled equations \eqref{eq:LLBar}
and \eqref{eq:2TM} forms a thermodynamically self-consistent non-stochastic
micromagnetic model. It is worth noting, that our model correctly
reproduces the specific heat of spins given by the mean-field approximation
in the temperature range of $[0,T_{c}]$. 

We would like to emphasize that neither LLB nor M3T nor the proposed
model account for the angular momentum conservation law and thereby
cannot be strictly used to identify the microscopic scattering mechanisms
responsible for the ultrafast heat-induced demagnetization. This can
only be achieved using full spin-electron-phonon model in the spirits
of the one used by \citet{Ma2012}. 

Now we apply our model to the problem of the ultrafast laser heating
of Nickel that was systematically investigated by \citet{Roth2012}.
Hereafter we refer to the data acquired at fixed ambient temperature
(of $T_{amb}=300\, K$) using varying laser fluence as \textit{fluence}
\textit{data}, while the data acquired at fixed laser fluence (of
$F_{0}=35\, Jm^{-2}$) using varying ambient temperature as the \textit{temperature
data}. In the simulations we neglect the direct magnon-phonon relaxation
mechanism, since in Nickel it happens on a timescale well beyond the
ultrafast dynamics\citep{Illg2013}. We also neglect any heat-transport,
i.e. $k_{e}=k_{p}\equiv0$, since (a) sample thickness is assumed
to be comparable to the skin-depth, (b) the ratio between the diameters
of the pump and probe spots used in the experiments is around $250:1$
and (c) the heat transfer is much slower than the local longitudinal
magnetization dynamics. The values of the Debye temperature $\theta_{D}=390\, K$,
$\gamma=4.51\cdot10^{-3}\, JK^{-2}mol^{-1}$ and $T_{C}=633\, K$
are all extracted from Ref. \citep{Meschter1981}. The atomic moment
of Nickel is $J=\frac{1}{2}.$ The laser pulse is assumed to be Gaussian
with FWHM $50\, fs$. The light absorption $A$ , $G_{e-p}$ and $\alpha_{\parallel}$
were all estimated using the Levenberg-Marquardt least-squares fitting
algorithm from the SciPy set of libraries\citep{Jones2001--}. The
solutions to the system of equations \eqref{eq:LLBar} and \eqref{eq:2TM}
are calulcated using the in-house developed open-source \textit{hotspin}
micromagnetic solver \citep{hotspin}. 

\begin{figure}
\includegraphics[width=8.5cm]{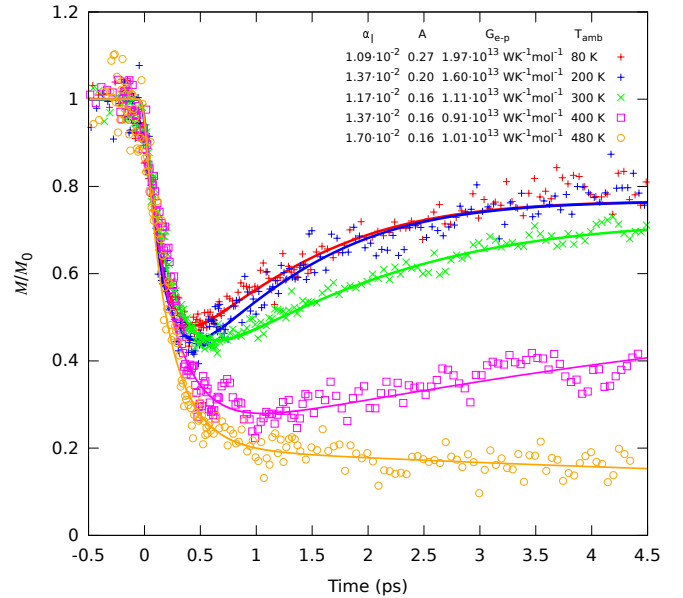}\caption{Temperature dependence of the longitudinal magnetization dynamics.
The solid lines represent the best fits of the proposed micromagnetic
model to the experimental data from Ref. \citep{Roth2012} (shown
as symbols). \label{fig:temperature-fit}}
\end{figure}

\begin{figure}
\includegraphics[width=8.5cm]{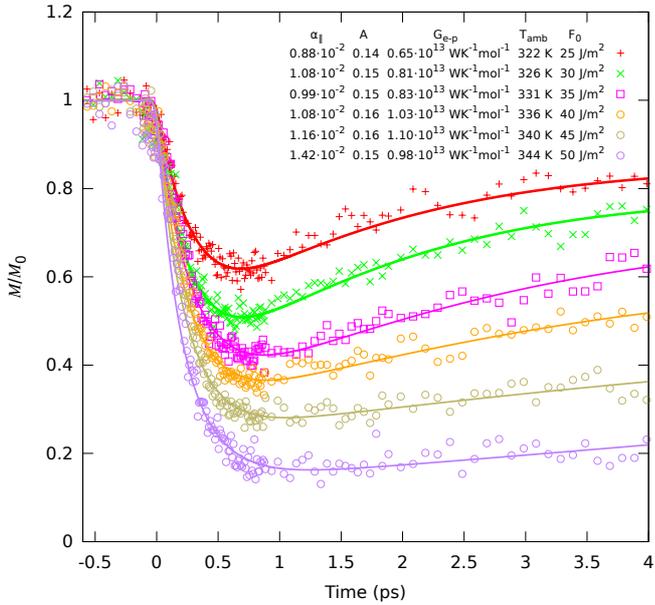}\caption{Fluence dependence of the longitudinal magnetization dynamics. The
solid lines represent the best fits of the proposed micromagnetic
model to the experimental data from Ref. \citep{Roth2012} (shown
as symbols).\label{fig:fluence-fit}}
\end{figure}

Our analysis shows that for the fluence data, the estimated values
of the aforementioned parameters were significantly different from
those estimated from the temperature data. In a private communication,
the authors of the experimental study confirmed that (a) the temperature
and fluence data were acquired from different samples and (b) the
fluence data acquisition was performed without any means of the temperature
control leading to the accumulation of the residual heat. We account
for this effect by introducing the \textit{ad-hoc} linear dependence
of the ambient temperature on the laser fluence, i.e. $T_{amb}=(8.89\, KJ^{-1}m^{2})F_{0}$.

\begin{figure}
\includegraphics[width=8.5cm]{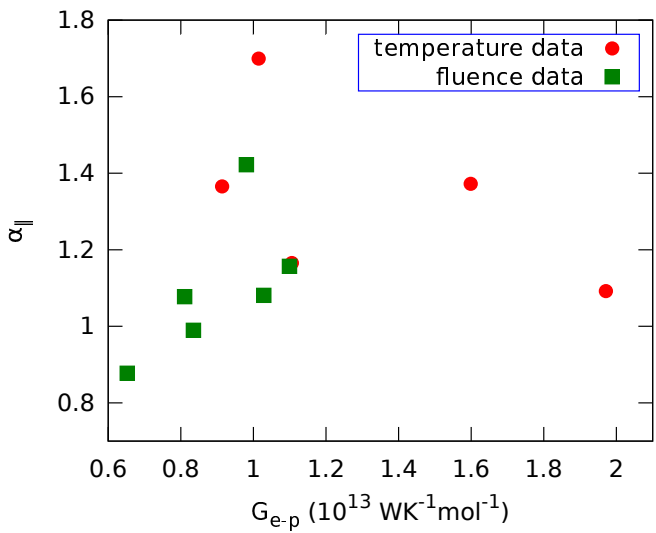}\caption{Zero-order relaxation constant with respect to the electron-phonon
coupling constant as extracted by fitting the proposed micromagnetic
model to the experimental data from Ref. \citep{Roth2012}. \label{fig:AlphaVSGep}}
\end{figure}

The best fits of our model to the experimental data are shown in Fig.
\ref{fig:temperature-fit} and Fig. \ref{fig:fluence-fit} for the
temperature and fluence data, respectively. The correlation between
the zero-order relativistic relaxation and electron-phonon coupling
constants is shown in Fig. \eqref{fig:AlphaVSGep}. The fit shows
almost no correlation between these two quantities suggesting that
the ultrafast angular momentum transfer happens only within the spin-electron
subsystem. This is in contrast to M3T model that assumes that the
angular momentum is transferred to the lattice and predicts a linear
dependence between (longitudinal) damping and electron-phonon coupling
constants. However, the importance of the electron-phonon coupling
in the ultrafast magnetization dynamics should not be underestimated,
since it provides the ultrafast energy dissipation channel (possibly
via the inelastic electron-phonon scattering) and, thereby, maintains
the magnetization recover. So our findings question the widely accepted
hypothesis of the phonon-mediated Elliot-Yafet mechanism dominance
in the ultrafast magnetization dynamics, consistent with \textit{ab-initio}
calculations performed by \citet{Carva2011} and partly consistent
with estimations of \citet{Illg2013} who suggested to include the
spin-electron scattering into consideration of the ultrafast magnetization
dynamics.

For the temperature data, a gradual decrease of the electron-phonon
coupling constant is observed, consistent with the electron DOS calculations
from Ref. \citep{Lin2007}. In contrast, for the fluence data the
opposite effect is observed. We believe this is artificial, since
the exact experimental conditions are unknown in this case. This observation
urges higher quality measurements with finer steps in both laser fluence
and ambient temperature.

The fit gives the following bounds for the relevant parameters: $\alpha_{\parallel}=(1.34\pm0.24)\cdot10^{-2}$
consistent with $\alpha_{LL}=1.30\cdot10^{-2}$ estimated from Q-band
FMR measurements \citep{Inaba2006}, $A=0.19\pm0.05$ consistent with
$A=0.21$ measured at $T_{amb}=4.2\, K$ using continuous excitation
of $1\,\mu m$ wavelength\citep{Biondi1968}. Unfortunately, there
is no consistent data on the value of the macroscopic electron-phonon
coupling constant, since it is typically estimated indirectly, e.
g. using the three-temperature model where thermodynamic parameters
are assumed to be constant \citep{Beaurepaire1996}. Nevertheless,
we estimated that $G_{e-p}(T_{e})\in[0.91,1.97]\cdot10^{13}\, WK^{-1}mol^{-1}$
that is below the upper bound of $G_{e-p}(0)=2.13\cdot10^{13}\, WK^{-1}mol^{-1}$
provided by the electron DOS calculations\citep{Lin2007}.

In contrast to the LLB and M3T models, we assume that the relaxation
constants are independent of the electron and (or) phonon temperatures.
In fact, the fitting shows a weak temperature dependence of the zero-order
relativistic damping constant for $T_{amb}\leqslant400\, K$, with
prominent enhancement for $T_{amb}=480\, K$. This might suggest an
activation (or significant enhancement) of an alternative relaxation
channel, e.g. phonon-mediated Elliot-Yafet mechanism or spin super-diffusion.
The fact that the value of the electron-phonon coupling constant is
also enhanced (contradictory to the electron DOS calculations) might
indeed be a sign of the inelastic phonon-mediated Elliot-Yafet mechanism
that pops-up only when exchange splitting becomes vanishing. The influence
of the non-local damping on the observed enhancement can only be estimated
using atomistic simulations.

In conclusion, we propose a model that solves two major problems of
finite-temperature micromagnetism: (i) it provides physical interpretation
of the relaxation terms and (ii) it fully complies with the energy
conservation law. We show that the model quantitatively reproduces
recent experimental data on the ultrafast magnetization dynamics in
Nickel. It could be readily used to (a) quantitatively estimate the
LLBar-specific relaxation constants from the experimental data and
(b) to explore applications of spin caloritronics and heat-assisted
magnetic recording. 

Authors would like to acknowledge Dr. Mirko Cinchetti for sharing
the experimental data and the details on its acquisition routines.

\bibliographystyle{apsrev4-1}
\bibliography{/run/media/mykola/LACIE2/BACKUP/PAPERS/references}

\begin{thebibliography}{36}%
\makeatletter
\providecommand \@ifxundefined [1]{%
 \@ifx{#1\undefined}
}%
\providecommand \@ifnum [1]{%
 \ifnum #1\expandafter \@firstoftwo
 \else \expandafter \@secondoftwo
 \fi
}%
\providecommand \@ifx [1]{%
 \ifx #1\expandafter \@firstoftwo
 \else \expandafter \@secondoftwo
 \fi
}%
\providecommand \natexlab [1]{#1}%
\providecommand \enquote  [1]{``#1''}%
\providecommand \bibnamefont  [1]{#1}%
\providecommand \bibfnamefont [1]{#1}%
\providecommand \citenamefont [1]{#1}%
\providecommand \href@noop [0]{\@secondoftwo}%
\providecommand \href [0]{\begingroup \@sanitize@url \@href}%
\providecommand \@href[1]{\@@startlink{#1}\@@href}%
\providecommand \@@href[1]{\endgroup#1\@@endlink}%
\providecommand \@sanitize@url [0]{\catcode `\\12\catcode `\$12\catcode
  `\&12\catcode `\#12\catcode `\^12\catcode `\_12\catcode `\%12\relax}%
\providecommand \@@startlink[1]{}%
\providecommand \@@endlink[0]{}%
\providecommand \url  [0]{\begingroup\@sanitize@url \@url }%
\providecommand \@url [1]{\endgroup\@href {#1}{\urlprefix }}%
\providecommand \urlprefix  [0]{URL }%
\providecommand \Eprint [0]{\href }%
\providecommand \doibase [0]{http://dx.doi.org/}%
\providecommand \selectlanguage [0]{\@gobble}%
\providecommand \bibinfo  [0]{\@secondoftwo}%
\providecommand \bibfield  [0]{\@secondoftwo}%
\providecommand \translation [1]{[#1]}%
\providecommand \BibitemOpen [0]{}%
\providecommand \bibitemStop [0]{}%
\providecommand \bibitemNoStop [0]{.\EOS\space}%
\providecommand \EOS [0]{\spacefactor3000\relax}%
\providecommand \BibitemShut  [1]{\csname bibitem#1\endcsname}%
\let\auto@bib@innerbib\@empty
%</preamble>
\bibitem [{\citenamefont {Landau}\ and\ \citenamefont
  {Lifshitz}(1935)}]{Landau1935}%
  \BibitemOpen
  \bibfield  {author} {\bibinfo {author} {\bibfnamefont {L.~D.}\ \bibnamefont
  {Landau}}\ and\ \bibinfo {author} {\bibfnamefont {E.}~\bibnamefont
  {Lifshitz}},\ }\href@noop {} {\bibfield  {journal} {\bibinfo  {journal}
  {Phys. Z. Sowjet.}\ }\textbf {\bibinfo {volume} {8}},\ \bibinfo {pages} {153}
  (\bibinfo {year} {1935})}\BibitemShut {NoStop}%
\bibitem [{\citenamefont {Beaurepaire}\ \emph {et~al.}(1996)\citenamefont
  {Beaurepaire}, \citenamefont {Merle}, \citenamefont {Daunois},\ and\
  \citenamefont {Bigot}}]{Beaurepaire1996}%
  \BibitemOpen
  \bibfield  {author} {\bibinfo {author} {\bibfnamefont {E.}~\bibnamefont
  {Beaurepaire}}, \bibinfo {author} {\bibfnamefont {J.-C.}\ \bibnamefont
  {Merle}}, \bibinfo {author} {\bibfnamefont {A.}~\bibnamefont {Daunois}}, \
  and\ \bibinfo {author} {\bibfnamefont {J.-Y.}\ \bibnamefont {Bigot}},\ }\href
  {http://link.aps.org/doi/10.1103/PhysRevLett.76.4250} {\bibfield  {journal}
  {\bibinfo  {journal} {Phys. Rev. Lett.}\ }\textbf {\bibinfo {volume} {76}},\
  \bibinfo {pages} {4250} (\bibinfo {year} {1996})}\BibitemShut {NoStop}%
\bibitem [{\citenamefont {Koopmans}\ \emph {et~al.}(2005)\citenamefont
  {Koopmans}, \citenamefont {Ruigrok}, \citenamefont {Longa},\ and\
  \citenamefont {de~Jonge}}]{Koopmans2005}%
  \BibitemOpen
  \bibfield  {author} {\bibinfo {author} {\bibfnamefont {B.}~\bibnamefont
  {Koopmans}}, \bibinfo {author} {\bibfnamefont {J.~J.~M.}\ \bibnamefont
  {Ruigrok}}, \bibinfo {author} {\bibfnamefont {F.~D.}\ \bibnamefont {Longa}},
  \ and\ \bibinfo {author} {\bibfnamefont {W.~J.~M.}\ \bibnamefont
  {de~Jonge}},\ }\href {\doibase 10.1103/PhysRevLett.95.267207} {\bibfield
  {journal} {\bibinfo  {journal} {Phys. Rev. Lett.}\ }\textbf {\bibinfo
  {volume} {95}},\ \bibinfo {pages} {267207} (\bibinfo {year}
  {2005})}\BibitemShut {NoStop}%
\bibitem [{\citenamefont {Radu}\ \emph {et~al.}(2009)\citenamefont {Radu},
  \citenamefont {Woltersdorf}, \citenamefont {Kiessling}, \citenamefont
  {Melnikov}, \citenamefont {Bovensiepen}, \citenamefont {Thiele},\ and\
  \citenamefont {Back}}]{Radu2009}%
  \BibitemOpen
  \bibfield  {author} {\bibinfo {author} {\bibfnamefont {I.}~\bibnamefont
  {Radu}}, \bibinfo {author} {\bibfnamefont {G.}~\bibnamefont {Woltersdorf}},
  \bibinfo {author} {\bibfnamefont {M.}~\bibnamefont {Kiessling}}, \bibinfo
  {author} {\bibfnamefont {A.}~\bibnamefont {Melnikov}}, \bibinfo {author}
  {\bibfnamefont {U.}~\bibnamefont {Bovensiepen}}, \bibinfo {author}
  {\bibfnamefont {J.-U.}\ \bibnamefont {Thiele}}, \ and\ \bibinfo {author}
  {\bibfnamefont {C.~H.}\ \bibnamefont {Back}},\ }\href {\doibase
  10.1103/PhysRevLett.102.117201} {\bibfield  {journal} {\bibinfo  {journal}
  {Phys. Rev. Lett.}\ }\textbf {\bibinfo {volume} {102}},\ \bibinfo {pages}
  {117201} (\bibinfo {year} {2009})}\BibitemShut {NoStop}%
\bibitem [{\citenamefont {Koopmans}\ \emph {et~al.}(2010)\citenamefont
  {Koopmans}, \citenamefont {Malinowski}, \citenamefont {Dalla~Longa},
  \citenamefont {Steiauf}, \citenamefont {Fähnle}, \citenamefont {Roth},
  \citenamefont {Cinchetti},\ and\ \citenamefont {Aeschlimann}}]{Koopmans2010}%
  \BibitemOpen
  \bibfield  {author} {\bibinfo {author} {\bibfnamefont {B.}~\bibnamefont
  {Koopmans}}, \bibinfo {author} {\bibfnamefont {G.}~\bibnamefont
  {Malinowski}}, \bibinfo {author} {\bibfnamefont {F.}~\bibnamefont
  {Dalla~Longa}}, \bibinfo {author} {\bibfnamefont {D.}~\bibnamefont
  {Steiauf}}, \bibinfo {author} {\bibfnamefont {M.}~\bibnamefont {Fähnle}},
  \bibinfo {author} {\bibfnamefont {T.}~\bibnamefont {Roth}}, \bibinfo {author}
  {\bibfnamefont {M.}~\bibnamefont {Cinchetti}}, \ and\ \bibinfo {author}
  {\bibfnamefont {M.}~\bibnamefont {Aeschlimann}},\ }\href {\doibase
  10.1038/nmat2593} {\bibfield  {journal} {\bibinfo  {journal} {Nat. Mater.}\
  }\textbf {\bibinfo {volume} {9}},\ \bibinfo {pages} {259} (\bibinfo {year}
  {2010})}\BibitemShut {NoStop}%
\bibitem [{\citenamefont {Zhang}\ and\ \citenamefont
  {Zhang}(2009)}]{Zhang2009}%
  \BibitemOpen
  \bibfield  {author} {\bibinfo {author} {\bibfnamefont {S.}~\bibnamefont
  {Zhang}}\ and\ \bibinfo {author} {\bibfnamefont {S.~S.-L.}\ \bibnamefont
  {Zhang}},\ }\href {\doibase 10.1103/PhysRevLett.102.086601} {\bibfield
  {journal} {\bibinfo  {journal} {Phys. Rev. Lett.}\ }\textbf {\bibinfo
  {volume} {102}},\ \bibinfo {pages} {086601} (\bibinfo {year}
  {2009})}\BibitemShut {NoStop}%
\bibitem [{\citenamefont {Nembach}\ \emph {et~al.}(2011)\citenamefont
  {Nembach}, \citenamefont {Silva}, \citenamefont {Shaw}, \citenamefont
  {Schneider}, \citenamefont {Carey}, \citenamefont {Maat},\ and\ \citenamefont
  {Childress}}]{Nembach2011}%
  \BibitemOpen
  \bibfield  {author} {\bibinfo {author} {\bibfnamefont {H.~T.}\ \bibnamefont
  {Nembach}}, \bibinfo {author} {\bibfnamefont {T.~J.}\ \bibnamefont {Silva}},
  \bibinfo {author} {\bibfnamefont {J.~M.}\ \bibnamefont {Shaw}}, \bibinfo
  {author} {\bibfnamefont {M.~L.}\ \bibnamefont {Schneider}}, \bibinfo {author}
  {\bibfnamefont {M.~J.}\ \bibnamefont {Carey}}, \bibinfo {author}
  {\bibfnamefont {S.}~\bibnamefont {Maat}}, \ and\ \bibinfo {author}
  {\bibfnamefont {J.~R.}\ \bibnamefont {Childress}},\ }\href {\doibase
  10.1103/PhysRevB.84.054424} {\bibfield  {journal} {\bibinfo  {journal} {Phys.
  Rev. B}\ }\textbf {\bibinfo {volume} {84}},\ \bibinfo {pages} {054424}
  (\bibinfo {year} {2011})}\BibitemShut {NoStop}%
\bibitem [{\citenamefont {Fähnle}\ and\ \citenamefont
  {Zhang}(2013)}]{Faehnle2013}%
  \BibitemOpen
  \bibfield  {author} {\bibinfo {author} {\bibfnamefont {M.}~\bibnamefont
  {Fähnle}}\ and\ \bibinfo {author} {\bibfnamefont {S.}~\bibnamefont
  {Zhang}},\ }\href {\doibase 10.1016/j.jmmm.2012.09.024} {\bibfield  {journal}
  {\bibinfo  {journal} {J. Magn. Magn. Mater.}\ }\textbf {\bibinfo {volume}
  {326}},\ \bibinfo {pages} {232} (\bibinfo {year} {2013})}\BibitemShut
  {NoStop}%
\bibitem [{\citenamefont {Turgut}\ \emph {et~al.}(2013)\citenamefont {Turgut},
  \citenamefont {La-o vorakiat}, \citenamefont {Shaw}, \citenamefont
  {Grychtol}, \citenamefont {Nembach}, \citenamefont {Rudolf}, \citenamefont
  {Adam}, \citenamefont {Aeschlimann}, \citenamefont {Schneider}, \citenamefont
  {Silva}, \citenamefont {Murnane}, \citenamefont {Kapteyn},\ and\
  \citenamefont {Mathias}}]{Turgut2013}%
  \BibitemOpen
  \bibfield  {author} {\bibinfo {author} {\bibfnamefont {E.}~\bibnamefont
  {Turgut}}, \bibinfo {author} {\bibfnamefont {C.}~\bibnamefont {La-o
  vorakiat}}, \bibinfo {author} {\bibfnamefont {J.~M.}\ \bibnamefont {Shaw}},
  \bibinfo {author} {\bibfnamefont {P.}~\bibnamefont {Grychtol}}, \bibinfo
  {author} {\bibfnamefont {H.~T.}\ \bibnamefont {Nembach}}, \bibinfo {author}
  {\bibfnamefont {D.}~\bibnamefont {Rudolf}}, \bibinfo {author} {\bibfnamefont
  {R.}~\bibnamefont {Adam}}, \bibinfo {author} {\bibfnamefont {M.}~\bibnamefont
  {Aeschlimann}}, \bibinfo {author} {\bibfnamefont {C.~M.}\ \bibnamefont
  {Schneider}}, \bibinfo {author} {\bibfnamefont {T.~J.}\ \bibnamefont
  {Silva}}, \bibinfo {author} {\bibfnamefont {M.~M.}\ \bibnamefont {Murnane}},
  \bibinfo {author} {\bibfnamefont {H.~C.}\ \bibnamefont {Kapteyn}}, \ and\
  \bibinfo {author} {\bibfnamefont {S.}~\bibnamefont {Mathias}},\ }\href
  {\doibase 10.1103/PhysRevLett.110.197201} {\bibfield  {journal} {\bibinfo
  {journal} {Phys. Rev. Lett.}\ }\textbf {\bibinfo {volume} {110}},\ \bibinfo
  {pages} {197201} (\bibinfo {year} {2013})}\BibitemShut {NoStop}%
\bibitem [{\citenamefont {Atxitia}\ \emph {et~al.}(2010)\citenamefont
  {Atxitia}, \citenamefont {Chubykalo-Fesenko}, \citenamefont {Walowski},
  \citenamefont {Mann},\ and\ \citenamefont {Münzenberg}}]{Atxitia2010}%
  \BibitemOpen
  \bibfield  {author} {\bibinfo {author} {\bibfnamefont {U.}~\bibnamefont
  {Atxitia}}, \bibinfo {author} {\bibfnamefont {O.}~\bibnamefont
  {Chubykalo-Fesenko}}, \bibinfo {author} {\bibfnamefont {J.}~\bibnamefont
  {Walowski}}, \bibinfo {author} {\bibfnamefont {A.}~\bibnamefont {Mann}}, \
  and\ \bibinfo {author} {\bibfnamefont {M.}~\bibnamefont {Münzenberg}},\
  }\href {\doibase 10.1103/PhysRevB.81.174401} {\bibfield  {journal} {\bibinfo
  {journal} {Phys. Rev. B}\ }\textbf {\bibinfo {volume} {81}},\ \bibinfo
  {pages} {174401} (\bibinfo {year} {2010})}\BibitemShut {NoStop}%
\bibitem [{\citenamefont {Atxitia}\ and\ \citenamefont
  {Chubykalo-Fesenko}(2011)}]{Atxitia2011}%
  \BibitemOpen
  \bibfield  {author} {\bibinfo {author} {\bibfnamefont {U.}~\bibnamefont
  {Atxitia}}\ and\ \bibinfo {author} {\bibfnamefont {O.}~\bibnamefont
  {Chubykalo-Fesenko}},\ }\href {\doibase 10.1103/PhysRevB.84.144414}
  {\bibfield  {journal} {\bibinfo  {journal} {Phys. Rev. B}\ }\textbf {\bibinfo
  {volume} {84}},\ \bibinfo {pages} {144414} (\bibinfo {year}
  {2011})}\BibitemShut {NoStop}%
\bibitem [{\citenamefont {Kaganov}\ \emph {et~al.}(1957)\citenamefont
  {Kaganov}, \citenamefont {Lifshitz},\ and\ \citenamefont
  {Tanatarov}}]{Kaganov1957}%
  \BibitemOpen
  \bibfield  {author} {\bibinfo {author} {\bibfnamefont {M.}~\bibnamefont
  {Kaganov}}, \bibinfo {author} {\bibfnamefont {I.}~\bibnamefont {Lifshitz}}, \
  and\ \bibinfo {author} {\bibfnamefont {L.}~\bibnamefont {Tanatarov}},\
  }\href@noop {} {\bibfield  {journal} {\bibinfo  {journal} {J. Exp. Theor.
  Phys.}\ }\textbf {\bibinfo {volume} {4}},\ \bibinfo {pages} {173} (\bibinfo
  {year} {1957})}\BibitemShut {NoStop}%
\bibitem [{\citenamefont {Garanin}(1997)}]{Garanin1997}%
  \BibitemOpen
  \bibfield  {author} {\bibinfo {author} {\bibfnamefont {D.~A.}\ \bibnamefont
  {Garanin}},\ }\href {\doibase 10.1103/PhysRevB.55.3050} {\bibfield  {journal}
  {\bibinfo  {journal} {Phys. Rev. B}\ }\textbf {\bibinfo {volume} {55}},\
  \bibinfo {pages} {3050} (\bibinfo {year} {1997})}\BibitemShut {NoStop}%
\bibitem [{\citenamefont {Mayergoyz}\ \emph {et~al.}(2012)\citenamefont
  {Mayergoyz}, \citenamefont {Bertotti}, \citenamefont {Serpico}, \citenamefont
  {Liu},\ and\ \citenamefont {Lee}}]{Mayergoyz2012}%
  \BibitemOpen
  \bibfield  {author} {\bibinfo {author} {\bibfnamefont {I.}~\bibnamefont
  {Mayergoyz}}, \bibinfo {author} {\bibfnamefont {G.}~\bibnamefont {Bertotti}},
  \bibinfo {author} {\bibfnamefont {C.}~\bibnamefont {Serpico}}, \bibinfo
  {author} {\bibfnamefont {Z.}~\bibnamefont {Liu}}, \ and\ \bibinfo {author}
  {\bibfnamefont {A.}~\bibnamefont {Lee}},\ }\href {\doibase 10.1063/1.3670510}
  {\bibfield  {journal} {\bibinfo  {journal} {Journal of Applied Physics}\
  }\textbf {\bibinfo {volume} {111}},\  (\bibinfo {year} {2012})}\BibitemShut
  {NoStop}%
\bibitem [{\citenamefont {Ma}\ \emph {et~al.}(2012)\citenamefont {Ma},
  \citenamefont {Dudarev},\ and\ \citenamefont {Woo}}]{Ma2012}%
  \BibitemOpen
  \bibfield  {author} {\bibinfo {author} {\bibfnamefont {P.-W.}\ \bibnamefont
  {Ma}}, \bibinfo {author} {\bibfnamefont {S.~L.}\ \bibnamefont {Dudarev}}, \
  and\ \bibinfo {author} {\bibfnamefont {C.~H.}\ \bibnamefont {Woo}},\ }\href
  {\doibase 10.1103/PhysRevB.85.184301} {\bibfield  {journal} {\bibinfo
  {journal} {Phys. Rev. B}\ }\textbf {\bibinfo {volume} {85}},\ \bibinfo
  {pages} {184301} (\bibinfo {year} {2012})}\BibitemShut {NoStop}%
\bibitem [{\citenamefont {Chimata}\ \emph {et~al.}(2012)\citenamefont
  {Chimata}, \citenamefont {Bergman}, \citenamefont {Bergqvist}, \citenamefont
  {Sanyal},\ and\ \citenamefont {Eriksson}}]{Chimata2012}%
  \BibitemOpen
  \bibfield  {author} {\bibinfo {author} {\bibfnamefont {R.}~\bibnamefont
  {Chimata}}, \bibinfo {author} {\bibfnamefont {A.}~\bibnamefont {Bergman}},
  \bibinfo {author} {\bibfnamefont {L.}~\bibnamefont {Bergqvist}}, \bibinfo
  {author} {\bibfnamefont {B.}~\bibnamefont {Sanyal}}, \ and\ \bibinfo {author}
  {\bibfnamefont {O.}~\bibnamefont {Eriksson}},\ }\href {\doibase
  10.1103/PhysRevLett.109.157201} {\bibfield  {journal} {\bibinfo  {journal}
  {Phys. Rev. Lett.}\ }\textbf {\bibinfo {volume} {109}},\ \bibinfo {pages}
  {157201} (\bibinfo {year} {2012})}\BibitemShut {NoStop}%
\bibitem [{\citenamefont {Bar'yakhtar}\ \emph {et~al.}(1986)\citenamefont
  {Bar'yakhtar}, \citenamefont {Ivanov}, \citenamefont {Soboleva},\ and\
  \citenamefont {Sukstanskii}}]{Baryakhtar1986}%
  \BibitemOpen
  \bibfield  {author} {\bibinfo {author} {\bibfnamefont {V.~G.}\ \bibnamefont
  {Bar'yakhtar}}, \bibinfo {author} {\bibfnamefont {B.~A.}\ \bibnamefont
  {Ivanov}}, \bibinfo {author} {\bibfnamefont {T.~K.}\ \bibnamefont
  {Soboleva}}, \ and\ \bibinfo {author} {\bibfnamefont {A.~L.}\ \bibnamefont
  {Sukstanskii}},\ }\href@noop {} {\bibfield  {journal} {\bibinfo  {journal}
  {Soviet Physics - JETP}\ }\textbf {\bibinfo {volume} {64}},\ \bibinfo {pages}
  {857} (\bibinfo {year} {1986})}\BibitemShut {NoStop}%
\bibitem [{\citenamefont {Baryakhtar}\ and\ \citenamefont
  {Baryakhtar}(1998)}]{Baryakhtar1998}%
  \BibitemOpen
  \bibfield  {author} {\bibinfo {author} {\bibfnamefont {I.~V.}\ \bibnamefont
  {Baryakhtar}}\ and\ \bibinfo {author} {\bibfnamefont {V.~G.}\ \bibnamefont
  {Baryakhtar}},\ }\href
  {http://scholar.google.com/scholar_host?q=info:Hl8VASosjvcJ:scholar.google.com/&output=viewport&pg=95}
  {\bibfield  {journal} {\bibinfo  {journal} {Ukr. J. Phys.}\ }\textbf
  {\bibinfo {volume} {43}},\ \bibinfo {pages} {1433} (\bibinfo {year}
  {1998})}\BibitemShut {NoStop}%
\bibitem [{\citenamefont {Bar'yakhtar}\ and\ \citenamefont
  {Danilevich}(2010)}]{Baryakhtar2010}%
  \BibitemOpen
  \bibfield  {author} {\bibinfo {author} {\bibfnamefont {V.~G.}\ \bibnamefont
  {Bar'yakhtar}}\ and\ \bibinfo {author} {\bibfnamefont {A.~G.}\ \bibnamefont
  {Danilevich}},\ }\href {\doibase 10.1063/1.3421029} {\bibfield  {journal}
  {\bibinfo  {journal} {Low Temp. Phys.}\ }\textbf {\bibinfo {volume} {36}},\
  \bibinfo {pages} {303} (\bibinfo {year} {2010})}\BibitemShut {NoStop}%
\bibitem [{\citenamefont {Sultan}\ \emph {et~al.}(2012)\citenamefont {Sultan},
  \citenamefont {Atxitia}, \citenamefont {Melnikov}, \citenamefont
  {Chubykalo-Fesenko},\ and\ \citenamefont {Bovensiepen}}]{Sultan2012}%
  \BibitemOpen
  \bibfield  {author} {\bibinfo {author} {\bibfnamefont {M.}~\bibnamefont
  {Sultan}}, \bibinfo {author} {\bibfnamefont {U.}~\bibnamefont {Atxitia}},
  \bibinfo {author} {\bibfnamefont {A.}~\bibnamefont {Melnikov}}, \bibinfo
  {author} {\bibfnamefont {O.}~\bibnamefont {Chubykalo-Fesenko}}, \ and\
  \bibinfo {author} {\bibfnamefont {U.}~\bibnamefont {Bovensiepen}},\ }\href
  {\doibase 10.1103/PhysRevB.85.184407} {\bibfield  {journal} {\bibinfo
  {journal} {Phys. Rev. B}\ }\textbf {\bibinfo {volume} {85}},\ \bibinfo
  {pages} {184407} (\bibinfo {year} {2012})}\BibitemShut {NoStop}%
\bibitem [{\citenamefont {Oguchi}(1955)}]{Oguchi1955}%
  \BibitemOpen
  \bibfield  {author} {\bibinfo {author} {\bibfnamefont {T.}~\bibnamefont
  {Oguchi}},\ }\href {\doibase 10.1143/PTP.13.148} {\bibfield  {journal}
  {\bibinfo  {journal} {Progress of Theoretical Physics}\ }\textbf {\bibinfo
  {volume} {13}},\ \bibinfo {pages} {148} (\bibinfo {year} {1955})}\BibitemShut
  {NoStop}%
\bibitem [{\citenamefont {Sparks}\ \emph {et~al.}(1961)\citenamefont {Sparks},
  \citenamefont {Loudon},\ and\ \citenamefont {Kittel}}]{Sparks1961}%
  \BibitemOpen
  \bibfield  {author} {\bibinfo {author} {\bibfnamefont {M.}~\bibnamefont
  {Sparks}}, \bibinfo {author} {\bibfnamefont {R.}~\bibnamefont {Loudon}}, \
  and\ \bibinfo {author} {\bibfnamefont {C.}~\bibnamefont {Kittel}},\ }\href
  {\doibase 10.1103/PhysRev.122.791} {\bibfield  {journal} {\bibinfo  {journal}
  {Phys. Rev.}\ }\textbf {\bibinfo {volume} {122}},\ \bibinfo {pages} {791}
  (\bibinfo {year} {1961})}\BibitemShut {NoStop}%
\bibitem [{\citenamefont {Pincus}\ \emph {et~al.}(1961)\citenamefont {Pincus},
  \citenamefont {Sparks},\ and\ \citenamefont {LeCraw}}]{Pincus1961}%
  \BibitemOpen
  \bibfield  {author} {\bibinfo {author} {\bibfnamefont {P.}~\bibnamefont
  {Pincus}}, \bibinfo {author} {\bibfnamefont {M.}~\bibnamefont {Sparks}}, \
  and\ \bibinfo {author} {\bibfnamefont {R.~C.}\ \bibnamefont {LeCraw}},\
  }\href {\doibase 10.1103/PhysRev.124.1015} {\bibfield  {journal} {\bibinfo
  {journal} {Phys. Rev.}\ }\textbf {\bibinfo {volume} {124}},\ \bibinfo {pages}
  {1015} (\bibinfo {year} {1961})}\BibitemShut {NoStop}%
\bibitem [{\citenamefont {Gurevich}\ and\ \citenamefont
  {Melkov}(1996)}]{A.G.Gurevich1996}%
  \BibitemOpen
  \bibfield  {author} {\bibinfo {author} {\bibfnamefont {A.~G.}\ \bibnamefont
  {Gurevich}}\ and\ \bibinfo {author} {\bibfnamefont {G.~A.}\ \bibnamefont
  {Melkov}},\ }\href {http://books.google.be/books?id=YgQtSvFIvFQC} {\emph
  {\bibinfo {title} {Magnetization Oscillations and Waves}}}\ (\bibinfo
  {publisher} {CRC Press},\ \bibinfo {address} {Boca Raton, FL},\ \bibinfo
  {year} {1996})\BibitemShut {NoStop}%
\bibitem [{\citenamefont {Overhauser}(1953)}]{Overhauser1953}%
  \BibitemOpen
  \bibfield  {author} {\bibinfo {author} {\bibfnamefont {A.~W.}\ \bibnamefont
  {Overhauser}},\ }\href {\doibase 10.1103/PhysRev.89.689} {\bibfield
  {journal} {\bibinfo  {journal} {Phys. Rev.}\ }\textbf {\bibinfo {volume}
  {89}},\ \bibinfo {pages} {689} (\bibinfo {year} {1953})}\BibitemShut
  {NoStop}%
\bibitem [{\citenamefont {Heine}(1967)}]{Heine1967}%
  \BibitemOpen
  \bibfield  {author} {\bibinfo {author} {\bibfnamefont {V.}~\bibnamefont
  {Heine}},\ }\href {\doibase 10.1103/PhysRev.153.673} {\bibfield  {journal}
  {\bibinfo  {journal} {Phys. Rev.}\ }\textbf {\bibinfo {volume} {153}},\
  \bibinfo {pages} {673} (\bibinfo {year} {1967})}\BibitemShut {NoStop}%
\bibitem [{\citenamefont {Majlis}(2007)}]{Majlis2007}%
  \BibitemOpen
  \bibfield  {author} {\bibinfo {author} {\bibfnamefont {N.}~\bibnamefont
  {Majlis}},\ }\href {http://books.google.be/books?id=EKT4YaovXuYC} {\emph
  {\bibinfo {title} {The Quantum Theory of Magnetism}}},\ \bibinfo {edition}
  {2nd}\ ed.\ (\bibinfo  {publisher} {World Scientific},\ \bibinfo {address}
  {Singapore},\ \bibinfo {year} {2007})\BibitemShut {NoStop}%
\bibitem [{\citenamefont {Roth}\ \emph {et~al.}(2012)\citenamefont {Roth},
  \citenamefont {Schellekens}, \citenamefont {Alebrand}, \citenamefont
  {Schmitt}, \citenamefont {Steil}, \citenamefont {Koopmans}, \citenamefont
  {Cinchetti},\ and\ \citenamefont {Aeschlimann}}]{Roth2012}%
  \BibitemOpen
  \bibfield  {author} {\bibinfo {author} {\bibfnamefont {T.}~\bibnamefont
  {Roth}}, \bibinfo {author} {\bibfnamefont {A.~J.}\ \bibnamefont
  {Schellekens}}, \bibinfo {author} {\bibfnamefont {S.}~\bibnamefont
  {Alebrand}}, \bibinfo {author} {\bibfnamefont {O.}~\bibnamefont {Schmitt}},
  \bibinfo {author} {\bibfnamefont {D.}~\bibnamefont {Steil}}, \bibinfo
  {author} {\bibfnamefont {B.}~\bibnamefont {Koopmans}}, \bibinfo {author}
  {\bibfnamefont {M.}~\bibnamefont {Cinchetti}}, \ and\ \bibinfo {author}
  {\bibfnamefont {M.}~\bibnamefont {Aeschlimann}},\ }\href {\doibase
  10.1103/PhysRevX.2.021006} {\bibfield  {journal} {\bibinfo  {journal} {Phys.
  Rev. X}\ }\textbf {\bibinfo {volume} {2}},\ \bibinfo {pages} {021006}
  (\bibinfo {year} {2012})}\BibitemShut {NoStop}%
\bibitem [{\citenamefont {Illg}\ \emph {et~al.}(2013)\citenamefont {Illg},
  \citenamefont {Haag},\ and\ \citenamefont {Fähnle}}]{Illg2013}%
  \BibitemOpen
  \bibfield  {author} {\bibinfo {author} {\bibfnamefont {C.}~\bibnamefont
  {Illg}}, \bibinfo {author} {\bibfnamefont {M.}~\bibnamefont {Haag}}, \ and\
  \bibinfo {author} {\bibfnamefont {M.}~\bibnamefont {Fähnle}},\ }\href
  {\doibase 10.1103/PhysRevB.88.214404} {\bibfield  {journal} {\bibinfo
  {journal} {Phys. Rev. B}\ }\textbf {\bibinfo {volume} {88}},\ \bibinfo
  {pages} {214404} (\bibinfo {year} {2013})}\BibitemShut {NoStop}%
\bibitem [{\citenamefont {Meschter}\ \emph {et~al.}(1981)\citenamefont
  {Meschter}, \citenamefont {Wright}, \citenamefont {Brooks},\ and\
  \citenamefont {Kollie}}]{Meschter1981}%
  \BibitemOpen
  \bibfield  {author} {\bibinfo {author} {\bibfnamefont {P.~J.}\ \bibnamefont
  {Meschter}}, \bibinfo {author} {\bibfnamefont {J.~W.}\ \bibnamefont
  {Wright}}, \bibinfo {author} {\bibfnamefont {C.~R.}\ \bibnamefont {Brooks}},
  \ and\ \bibinfo {author} {\bibfnamefont {T.~G.}\ \bibnamefont {Kollie}},\
  }\href {http://www.sciencedirect.com/science/article/pii/0022369781901748}
  {\bibfield  {journal} {\bibinfo  {journal} {Journal of Physics and Chemistry
  of Solids}\ }\textbf {\bibinfo {volume} {42}},\ \bibinfo {pages} {861}
  (\bibinfo {year} {1981})}\BibitemShut {NoStop}%
\bibitem [{\citenamefont {Jones}\ \emph {et~al.}(01  )\citenamefont {Jones},
  \citenamefont {Oliphant}, \citenamefont {Peterson} \emph
  {et~al.}}]{Jones2001--}%
  \BibitemOpen
  \bibfield  {author} {\bibinfo {author} {\bibfnamefont {E.}~\bibnamefont
  {Jones}}, \bibinfo {author} {\bibfnamefont {T.}~\bibnamefont {Oliphant}},
  \bibinfo {author} {\bibfnamefont {P.}~\bibnamefont {Peterson}},  \emph
  {et~al.},\ }\href {http://www.scipy.org/} {\enquote {\bibinfo {title}
  {{SciPy}: Open source scientific tools for {Python}},}\ } (\bibinfo {year}
  {2001--})\BibitemShut {NoStop}%
\bibitem [{\citenamefont {Dvornik}(2014)}]{hotspin}%
  \BibitemOpen
  \bibfield  {author} {\bibinfo {author} {\bibfnamefont {M.}~\bibnamefont
  {Dvornik}},\ }\href@noop {} {\enquote {\bibinfo {title} {\emph{hotspin} - a
  high performance finite-difference micromagnetic solver for
  finite-temperature simulations},}\ }\bibinfo {howpublished}
  {\url{https://github.com/godsic/hotspin}} (\bibinfo {year}
  {2014})\BibitemShut {NoStop}%
\bibitem [{\citenamefont {Carva}\ \emph {et~al.}(2011)\citenamefont {Carva},
  \citenamefont {Battiato},\ and\ \citenamefont {Oppeneer}}]{Carva2011}%
  \BibitemOpen
  \bibfield  {author} {\bibinfo {author} {\bibfnamefont {K.}~\bibnamefont
  {Carva}}, \bibinfo {author} {\bibfnamefont {M.}~\bibnamefont {Battiato}}, \
  and\ \bibinfo {author} {\bibfnamefont {P.~M.}\ \bibnamefont {Oppeneer}},\
  }\href {http://link.aps.org/doi/10.1103/PhysRevLett.107.207201} {\bibfield
  {journal} {\bibinfo  {journal} {Phys. Rev. Lett.}\ }\textbf {\bibinfo
  {volume} {107}},\ \bibinfo {pages} {207201} (\bibinfo {year}
  {2011})}\BibitemShut {NoStop}%
\bibitem [{\citenamefont {Lin}\ and\ \citenamefont {Zhigilei}(2007)}]{Lin2007}%
  \BibitemOpen
  \bibfield  {author} {\bibinfo {author} {\bibfnamefont {Z.}~\bibnamefont
  {Lin}}\ and\ \bibinfo {author} {\bibfnamefont {L.~V.}\ \bibnamefont
  {Zhigilei}},\ }\bibfield  {booktitle} {\emph {\bibinfo {booktitle}
  {Proceedings of the Fifth International Conference on Photo-Excited Processes
  and Applications (5-ICPEPA)}},\ }\href
  {http://www.sciencedirect.com/science/article/pii/S0169433207000815}
  {\bibfield  {journal} {\bibinfo  {journal} {Applied Surface Science}\
  }\textbf {\bibinfo {volume} {253}},\ \bibinfo {pages} {6295} (\bibinfo {year}
  {2007})}\BibitemShut {NoStop}%
\bibitem [{\citenamefont {Inaba}\ \emph {et~al.}(2006)\citenamefont {Inaba},
  \citenamefont {Asanuma}, \citenamefont {Igarashi}, \citenamefont {Mori},
  \citenamefont {Kirino}, \citenamefont {Koike},\ and\ \citenamefont
  {Morita}}]{Inaba2006}%
  \BibitemOpen
  \bibfield  {author} {\bibinfo {author} {\bibfnamefont {N.}~\bibnamefont
  {Inaba}}, \bibinfo {author} {\bibfnamefont {H.}~\bibnamefont {Asanuma}},
  \bibinfo {author} {\bibfnamefont {S.}~\bibnamefont {Igarashi}}, \bibinfo
  {author} {\bibfnamefont {S.}~\bibnamefont {Mori}}, \bibinfo {author}
  {\bibfnamefont {F.}~\bibnamefont {Kirino}}, \bibinfo {author} {\bibfnamefont
  {K.}~\bibnamefont {Koike}}, \ and\ \bibinfo {author} {\bibfnamefont
  {H.}~\bibnamefont {Morita}},\ }\bibfield  {booktitle} {\emph {\bibinfo
  {booktitle} {Magnetics, IEEE Transactions on}},\ }\href {\doibase
  10.1109/TMAG.2006.878813} {\bibfield  {journal} {\bibinfo  {journal}
  {Magnetics, IEEE Transactions on}\ }\textbf {\bibinfo {volume} {42}},\
  \bibinfo {pages} {2372} (\bibinfo {year} {2006})}\BibitemShut {NoStop}%
\bibitem [{\citenamefont {Biondi}\ and\ \citenamefont
  {Guobadia}(1968)}]{Biondi1968}%
  \BibitemOpen
  \bibfield  {author} {\bibinfo {author} {\bibfnamefont {M.~A.}\ \bibnamefont
  {Biondi}}\ and\ \bibinfo {author} {\bibfnamefont {A.~I.}\ \bibnamefont
  {Guobadia}},\ }\href {\doibase 10.1103/PhysRev.166.667} {\bibfield  {journal}
  {\bibinfo  {journal} {Phys. Rev.}\ }\textbf {\bibinfo {volume} {166}},\
  \bibinfo {pages} {667} (\bibinfo {year} {1968})}\BibitemShut {NoStop}%
\end{thebibliography}%
 
\end{document}